\documentclass[useAMS,usenatbib,aas_macros]{mn2e}

\usepackage{graphicx,graphics}
\usepackage{amsmath}

\title[Ejection of OB stars from very young star clusters]{Complete ejection of OB stars from very young star clusters and the formation of multiple populations}
\author[Long Wang et al.]{Long Wang$^{1,3,4}$\thanks{E-mail: longw@uni-bonn.de; }, Pavel Kroupa$^{1,2}$ and Tereza Jerabkova$^{1,2,5}$\\
$^{1}$ Helmholtz-Institut f\"ur Strahlen- und Kernphysik, University of Bonn, Nussallee 14-16, D-53115 Bonn, Germany \\
$^{2}$ Charles University in Prague, Faculty of Mathematics and Physics, Astronomical Institute,\\
V Hole\v{s}ovi\v{c}k\v{a}ch 2, CZ-180 00, Praha 8, Czech Republic \\
$^{3}$ Argelander Institut F\"ur Astronomie, Auf Dem H\"ugel 71, 53121, Bonn, Germany \\
$^{4}$ RIKEN Center for Computational Science, 7-1-26 Minatojima-minami-machi, Chuo-ku, Kobe, Hyogo 650-0047, Japan\\
$^{5}$ European Southern Observatory, Karl-Schwarzschild-Str. 2, 85748 Garching, Gernamy
}

\begin{document}

\date{Accepted --.  Received --; in original form --}

\pagerange{\pageref{firstpage}--\pageref{lastpage}} \pubyear{2002}

\maketitle

\label{firstpage}

\begin{abstract}

  Recently, three stellar sequences separated in age by about $1$~Myr were discovered in the ONC \citep{Beccari2017}.
  \cite{Kroupa2018} suggest that such small dense sub-populations eject all their OB stars via the decay of unstable few-body systems such that the gas can recombine and form new stars.
  This explains the multi-sequence phenomenon without introducing an extra mechanism into star formation theory.
  In this work, we apply the recently updated primordial binary distribution model \citep{Belloni2017} (implemented here in a new version of \textsc{mcluster}) and perform a large set of direct $N$-body simulations to investigate the feasibility of this dynamical scenario.
  Our results suggest that if $3$-$4$ OB stars in the ONC formed primordially mass-segregated in the cluster center with a maximum separation of about $0.003$~pc, all OB stars have a high chance ($\approx 50-70 \%$) to escape from the center and do not come back within $1$~Myr and
   the dynamical ejection scenario is a viable channel to form short-age-interval multi-population sequences as observed in the ONC. 
  This is also consistent with self-regulated star formation.
\end{abstract}


\section{Introduction}


The Orion Nebula Cluster (ONC) is the nearest star formation region where OB stars formed very recently.
It provides useful information to constrain the theory of star formation, the initial mass function (IMF) and primordial binary properties \cite[e.g.][]{Genzel1989}.

\cite{DR2010} compile an optical catalogue with a HR diagram for the ONC in the stellar mass range (0.08 - 8 M$_\odot$) and \cite{DR2010} and \cite{Jeffries2011} indicate the age spread of the PMS in the ONC to be 3-4 Myr.
\cite{Beccari2017} discovered three separated sequences in the optical color-magnitude diagram.
They suggest two possible explanations: unresolved binaries/multiples and three discrete episodes of star formation.
However, in the former scenario, the obvious two sequences in the color distribution require an abnormal mass ratio distribution of binaries, disfavouring the first possibility.
\cite{Kroupa2018} (here after ``ejection scenario'') provide a model to explain the second scenario via the stellar-dynamical ejection of all OB stars such that a new population of stars can form from the recombining gas (the ``repeated stellar-dynamical termination of feedback-halted filament-accretion model'')\footnote{A movie showing the scenario from our $N$-body simulations can be found in the Youtube link: \url{https://youtu.be/S4AarHAi_bA} }.

In the ejection scenario, the first episode of star formation occurs within $1$~Myr.
By optimally sampling the canonical IMF \citep{Kroupa2013,Yan2017}, for a cluster with $400$~M$_\odot$ in stars, four OB stars with mass $>8 $~M$\odot$ can form.
After the formation of the OB stars, the radiation feedback sufficiently suppresses the star formation in a very short time interval, which generates the first drop in the age distribution of the stars.
Then the dynamical encounters of the OB stars (two binary systems) can result in the ejections of the OB stars.
Indeed, OB runaways with a high velocity ($\ge 30$~km~s$^{-1}$) away from star-formation regions are observed \citep{Blaauw1961,Stone1979}.
$N$-body simulations have confirmed the possibility of OB runaway formation via strong interactions in few-body systems in ONC-like clusters \citep{Gualandris2004,Pflamm2006,Oh2015}.
If all OB stars are ejected within about $1$~Myr, the resumed infall of gas triggers the second episode of star formation.
This scenario can repeat to generate the third sequence.
Essentially, this scenario is a version of the gas-accretion model proposed by \cite{Pflamm2009}, according to which a star cluster is likely to accrete molecular gas once its ionizing stars have died.

However, as the scenario requires the strict condition that all OB stars should be ejected within about $1$~Myr, and this twice at least.
The suggestion that this may be possible is based on available $N$-body results without detailed calculations and numerical proof.
In this work, we study this scenario in detail, with self-consistent $N$-body simulations.

In Section ~\ref{sec:method}, we describe our $N$-body methods and the initial conditions of the models.
Section~\ref{sec:result} show the probabilities of OB ejections.
Then we discuss the ejection scenario with our simulation results in Section~\ref{sec:discussion}.
Finally, we draw conclusions in Section~\ref{sec:conclusion}.

\section{Methods}
\label{sec:method}

\subsection{OB ejection model}

To form the OB runaways, first the OB stars need to be in binaries which then encounter other massive stars or binaries.
If an unstable $3/4$-body system forms, the instability results in the decay of the system.
The binding energy of the binaries is transferred to the center-of-mass/stellar kinetic energy, and the binaries/stars can be ejected with high velocities.
Although the final fate of one specific chaotic system is difficult to predict, many $N$-body simulations provide a statistical basis for this model.
\cite{Pflamm2006} study this scenario for the Trapezium cluster.
For the specific set of Trapezium OB stars, they find the chance of the maximum number of stars being two with a pairwise distance below $0.05$~pc to be about $80\%$ after $1.0$~Myr, which indicates that an interruption of star formation for about $1$~Myr might occur.

When the OB stars form, they ionize the surrounding gas and heat it up, which suppresses the star formation.
The influence region of OB star ionization is a complex issue.
By a rough calculation, the ejection scenario suggests that a few $0.1$~Myr after formation of the OB stars, the star formation within $0.1-1.0$~pc from the OB stars is suppressed.
Thus, once all OB stars form, they need to form an unstable few-body system to be ejected out of a certain radius within about $0.5$~Myr.
Then the star formation can continue from the still infalling recombining gas.
These all-OB-ejection cases should happen twice in consecution to explain the three sequences in the ONC.

\subsection{N-body models}

\subsubsection{Direct $N$-body code}

In this work, we use the direct $N$-body code \textsc{nbody6++gpu} to perform the simulations \citep{Wang2015}.
This is a parallellization-optimized version of the state-of-the-art \textsc{nbody6} code \citep{Aarseth2003,Nitadori2012}.
\textsc{nbody6} is a fully self-consistent $N$-body code specifically designed for computing the evolution of star clusters, where collisional dynamics plays an important role due to the short two-body relaxation time.
The regularization algorithms used in \textsc{nbody6(++gpu)} \citep{KS1965,Mikkola1999} ensure an acceptable accuracy in the treatment of few-body interactions.
This is the crucial process that determines the ejection rate of OB stars.
\textsc{nbody6++gpu} uses the single and binary stellar evolution recipes (SSE/BSE) from \cite{Hurley2000,Hurley2002} \footnote{The current version of \textsc{nbody6} has the updated stellar evolution package (SSE/BSE) for stellar winds of massive stars and compact object formation. \textsc{nbody6++gpu} still keeps the original SSE/BSE version (before 2016). For this work, we only simulate the first two Myr of the star clusters. Thus this difference is not important. We use the version of \textsc{nbody6++gpu} on Github: https://github.com/nbodyx/Nbody6ppGPU (commits on Apr 27, 2018.)}.

As indicated in \cite{PZ2014}, the energy error due to the limited accuracy of the integration algorithm and digital limits inherent to the numerical computation can result in a significant divergence of the result for one specific chaotic system.
However, the statistical results of a large number of simulations ($10^3$) provide a similar frequency of remaining OB stars for different relative energy error.
This was also checked in \cite{Pflamm2006} for the energy error ranging from $10^{-2}$ to $10^{-12}$.
Thus in this work, we set the maximum relative energy error criterion to be $1/N$, (e.g., $5\times10^{-3}-5\times10^{-4}$ in our models).
All models which cannot satisfy the criterion are excluded \footnote{About $5\%$ sub-models of each major model are rejected due to the violation of energy conservation.}.

\subsubsection{Initial conditions}
\label{sec:init}

We provide two sets of models to study the OB ejections from young star clusters.
The first set is for the general case of young star clusters with different masses and the second is specifically designed for the ONC.
We use the newly updated version of \textsc{mcluster} \citep[originally from ][]{Kupper2011} to generate the initial models (see Appendix) \footnote{The updated version of the code can be found in Github: https://github.com/lwang-astro/mcluster (commits on May 15, 2018). Original: https://github.com/ahwkuepper/mcluster.}.

{\bf First set:}

In the Model Set 1, we sample the initial positions and velocities of stars in our $N$-body models from the Plummer density profile, since it is the simplest analytic solution of the collisionless Boltzmann equation with only two free parameters that can well describe the density structure of star clusters (originally for globular clusters, \citealp{Plummer1911}).
Young star clusters with an age less than a few Myr contain a significant fraction of gas.
In this work, we exclude the gas since it is not clear how the gas density is distributed in young star clusters, especially when considering the connection to the large-scale structure of gas in the star formation region including gas in-fall.
Also the complexity of gas dynamics and feedback introduces extra effects and uncertainties on determining the initial conditions, which are not the major physical processes focused on in this work.
A Solar-neighborhood tidal field is applied to determine the tidal radii of the clusters,
\begin{equation}
  R_{\mathrm t} = \left[ \frac{G M_{\mathrm{ecl}}}{4 A(A-B)} \right]^{1/3} ,
  \label{eq:tt}
\end{equation}
where $A$ and $B$ are Oort's constant with $A=14.4$ and $B=-12.0$~km~s$^{-1}$~kpc$^{-1}$ \cite{Aarseth2003}.
The escape criterion is set by the star exceeding twice the tidal radius.

We apply the canonical IMF from \cite{Kroupa2001} with optimal sampling \citep{Kroupa2013,Yan2017}.
For young open clusters, the total stellar mass ranges from $10^2$ to $10^3$~M$_\odot$.
Random sampling of the IMF results in large fluctuations of the masses and numbers of OB stars (see Section~\ref{sec:dis:2}).
However, \cite{Weidner2004,Weidner2006} suggest that observed very young clusters follow the maximum stellar mass, $m_{\mathrm{max}}$, cluster total stellar mass, $M_{\mathrm{ecl}}$, relation, and sorted sampling from the IMF with this relation provides a better fit to observational data than random sampling.
The later study by \cite{Kroupa2013,Weidner2013} suggests that star formation in embedded star clusters might be highly self-regulated since the observational scatter of IMF slopes is less than the Poisson noise.
Thus optimal sampling is recommended, in which the sampling from the IMF is not a random process and from which $m_{\mathrm{max}}$ follows, given an $M_{\mathrm{ecl}}$ \citep[see also][]{Weidner2014,Stephens2017}.
Although it is still under debate whether the optimal sampling from the IMF is preferred in nature, it provides an advantage for the purpose of this work.
When $M_{\mathrm{ecl}}$ is given, the number and masses of OB stars are fixed.
This makes the statistical studies of OB runaways more controlled.
Intriguingly, the ejection scenario suggests that the set of massive stars resulting from the optimal sampling may be consistent with the observed set.
We return to this issue in Section~\ref{sec:dis:2}.

We use an initially $100\%$ binary fraction with the \cite{Kroupa1995a,Kroupa1995b,Belloni2017} model to generate the binary parameters.
\cite{Kroupa1995a,Kroupa1995b} assumes that stars with mass less than a few $M_\odot$ are born with a universal binary distribution function and that the binary fraction is initially $100\%$ in star clusters,
and use inverse dynamical population synthesis to construct the initial binary period and mass-ratio distribution.
\cite{Belloni2017} provides an updated version of pre-main-sequence eigenevolution introduced by \cite{Kroupa1995b} to account for the correlation between mass-ratio, period and eccentricity for short-period systems.
This update (implemented in the new version of \textsc{mcluster}) is based on the observational constraint from the stellar color distribution of globular clusters.
For the high-mass (OB star) binaries ($>5$~M$_\odot$), we adopt the \cite{Sana2012} distribution, which is derived from O star samples of six nearby Galactic open clusters.
The distribution functions are based on \cite{Oh2015,Belloni2017}:
\begin{itemize}
\item Period distribution: $f_P = 0.23 \times (\log_{10}{P})^{-0.55}$;
\item Mass ratio distribution: uniform between $0.1$ to $0.9$;
\item Eccentricity distribution: $f_e = 0.55 \times e^{-0.45}$.
\end{itemize}

Based on the \cite{Kroupa1995a,Kroupa1995b} primordial-binary model and an analytical evolution model of binary properties \citep{Marks2011}, \cite{Marks2012} analyze the observational data of young star clusters and old open clusters and discover a weak cluster total mass - half-mass radius, $R_{\mathrm{h,0}}$, relation:
\begin{equation}
  R_{\mathrm{h,0}}/pc = 0.1^{+0.07}_{\mathrm{-0.04}} (M_{\mathrm{ecl}}/M_{\odot})^{0.13\pm0.04}.
\end{equation}
We adopt this relation to determine the $R_{\mathrm{h,0}}$ for our models.

By applying the above initial configuration, there are only two free parameters that can be adjusted: the total stellar mass $M_{\mathrm{ecl}}$ and the initial mass-segregation degree.
For general propose, we perform $10$ models with different mean $\langle M_{\mathrm{ecl}}\rangle$ ranging from $100$ to $1000$~M$_\odot$ with an interval of $100$~M$_\odot$\footnote{$\langle M_{\mathrm{ecl}}\rangle$ is used for generating stellar masses by optimal sampling from the IMF. Later, during the pre-main-sequence eigenevolution process \citep{Kroupa1995b,Belloni2017}, some of the secondary stars in binaries gain masses. Thus the final total cluster masses are slightly larger than $\langle M_{\mathrm{ecl}}\rangle$.}, which represent the typical mass range of young open clusters.
Such low mass clusters have numbers of stars $10^2$-$10^3$ and contain $1$-$10$ OB stars.
Thus chaotic dynamical effects are strong, and a large group of models is necessary to obtain well defined average quantities.
For each $M_{\mathrm{ecl}}$, we create $2001$ sub-models with the same initial conditions, except for randomly sampling the positions and velocities of single stars and binaries, and a slight modification of $M_{\mathrm{ecl}}$ (to include slight changes of OB star masses).
The $2001$ sub-models thus have masses $\langle M_{\mathrm{ecl}}\rangle \pm \sqrt{\langle M_{\mathrm{ecl}}\rangle} $~M$_\odot$ with equal mass intervals.
Due to the small differences of total masses, the corresponding initial number of OB stars, $N_{\mathrm{OB,0}}$, have a small variation for low-mass models.
Table~\ref{tab:sets} lists the $\langle M_{\mathrm{ecl}} \rangle$ values of these $10$ models and their corresponding total number of stars $N$, initial number of OB stars $N_{\mathrm{OB,0}}$, maximum stellar mass $m_{\mathrm{max}}$, $R_{\mathrm{h,0}}$, initial tidal radius $R_{\mathrm{t,0}}$ and initial number of OB stars ($m>8$~M$_\odot$).
We use two extreme mass-segregation cases: not-segregated, $S0$, and fully-segregated, $S1$, \citep{Baumgardt2008}.

{\bf Second set:}

In the second set, we focus on the conditions of the ONC cluster based on the \cite{Kroupa2018}.
This is designed to investigate which condition can result in a high chance of ejecting all OB stars.
The models are selected based on the results of the first set (see Section~\ref{sec:res:set1}).
We choose the M4-S1 set, where $3$ ($15\%$) or $4$ ($85\%$) OB stars concentrate in the cluster center.
In addition, $400$~M$_\odot$ is the approximate mass of one episode of star formation in the ONC \citep{Beccari2017}.
The $3-4$ OB stars are also a good configuration that can result in single/binary-binary 3/4-body interaction and eject each other out of the cluster.

First, we apply the parameters of the M4-S1 model to generate $2001$ models using the same procedure as in the first set.
The dynamical timescale of forming unstable multi-body OB star systems depends on the distance between the OB stars.
Thus we consider this as the dominate factor influencing the OB star ejection rate, and use Eq.~\ref{eq:scale} to scale the position and velocities of the center-of-masses of the binaries which include stars with mass $>8$~M$_\odot$,
\begin{equation}
  \label{eq:scale}
  \begin{aligned}
    \mathbf{r}_{\mathrm{s}} & =& \kappa \mathbf{r} ,\\
    \mathbf{v}_{\mathrm{s}} & =& \sqrt{\kappa} \mathbf{v} .\\
  \end{aligned}
\end{equation}
where $\mathbf{r}$ and $\mathbf{v}$ are the original coordinate and velocity vectors, and $\mathbf{r}_{\mathrm{s}}$ and $\mathbf{v}_{\mathrm{s}}$ are the scaled coordinate and velocity vectors.
Eq.~\ref{eq:scale} ensures the orbital kinetic energy and potential energy of the OB stars keep a constant ratio.
The values of $\kappa$ and corresponding model names are listed in Table~\ref{tab:sets} and these models are referred to as scaled Plummer models.

To modify the OB star density, we can also apply different models of distribution functions instead of the Plummer model.
\cite{King1966} provides a self-consistent density distribution model that is close to the singular isothermal sphere, but has a finite central density (core structure) and finite total mass (a radial cutoff).
The singular isothermal sphere has the density profile $\rho(r) \propto r^{-2}$.
By modifying the central concentration parameter $W_{\mathrm 0}$, we can construct models with different central density.
The King model with $W_{\mathrm 0}=6$ is close to the Plummer model \citep{Kroupa2008}.
In principle, when $W_{\mathrm 0}\Longrightarrow \infty$,  the profile approaches the singular isothermal profile .
However, in the case of M4-S1, only $3-4$ OB stars (in binaries) exist.
For $W_{\mathrm 0}>10$, statistically there is no obvious dependence of the density profile of the core region on $W_{\mathrm 0}$.
Thus $W_{\mathrm 0}=10$ represents the most dense core that can be sampled from the King model.
Here we add this King model (W10) for comparison with the scaled Plummer models.

Usually the King model cutoff radius is used as tidal radius, especially in observations.
But in our W10 model, we use the tidal radius $R_{\mathrm t}$ determined by Eq.~\ref{eq:tt}.
Thus the King model cutoff radius does not match the initial tidal radius, $R_{\mathrm t,0}$.

\begin{table*}
 \centering
 \begin{minipage}{140mm}
   \caption{Initial parameters of the two model sets. For Set 1, the mean total cluster stellar mass $\langle M_{\mathrm{ecl}} \rangle$; corresponding maximum stellar mass $m_{\mathrm{max}}$; average total number of stars $\langle N\rangle$ (binaries are resolved); total number of OB stars $N_{\mathrm{OB,0}}$; average (all sub-models for the non-segregated S0 and fully-segregated S1) initial half-mass radius $\langle R_{\mathrm{h,0}} \rangle$ and initial tidal radius $\langle R_{\mathrm{t,0}} \rangle$ are shown.
     The S0 and S1 models with the same mass have about $0.01$~pc difference in $R_{\mathrm{h,0}}$.  
     For the scaled Plummer models, the OB binary orbital scaling factor $\kappa$ is shown.
     The last column is the King model with $W=10$.}
   \label{tab:sets}
   \begin{tabular}{@{}lllllllllll@{}}
     \hline
     \multicolumn{11}{c}{Set 1}\\
     \hline
     Model & M1 & M2 & M3 & M4 & M5 & M6 & M7 & M8 & M9 & M10 \\
     $\langle M_{\mathrm{ecl}} \rangle$ [M$_\odot$] & 100 & 200 & 300 & 400 & 500 & 600 & 700 & 800 & 900 & 1000 \\
     $\langle N \rangle$ & 213 & 404 & 590 & 773 & 955 & 1135 & 1313 & 1491 & 1668 & 1845 \\
     $N_{\mathrm{OB,0}}$ & 1 & 1-2 & 2-3 & 3-4& 5 & 6 & 7 & 8 & 9 & 10 \\
     $m_{\mathrm{max}}$ [M$_\odot$] & 9.12 & 14.47 & 18.91 & 22.82 & 26.35 & 29.59 & 32.60 & 35.41 & 38.05 & 40.54 \\
     $\langle R_{\mathrm{h,0}} \rangle$ [pc] & 0.18 & 0.20 & 0.21 & 0.22 & 0.22 & 0.23 & 0.23 & 0.24 & 0.24 & 0.24 \\
     $\langle R_{\mathrm{t,0}} \rangle$ [pc] & 6.66 & 8.39 & 9.60 & 10.57 & 11.38 & 12.09 & 12.73 & 13.31 & 13.84 & 14.33 \\
     \hline
     \multicolumn{11}{c}{Set 2}\\
     \hline
     Model & K1 & K2 & K4 & K8 & K16 & K32 & \multicolumn{4}{l}{W10}  \\
     $\kappa$ & 1 & 1/2 & 1/4 & 1/8 & 1/16 & 1/32 & \multicolumn{4}{l}{King $W_{\mathrm 0}=10$}\\ 
     \hline
  \end{tabular}
\end{minipage}
\end{table*}

\section{Results}
\label{sec:result}

\subsection{Different $M_{\mathrm{ecl}}$}
\label{sec:res:set1}

\begin{figure*}
 \includegraphics[width=1.0\textwidth]{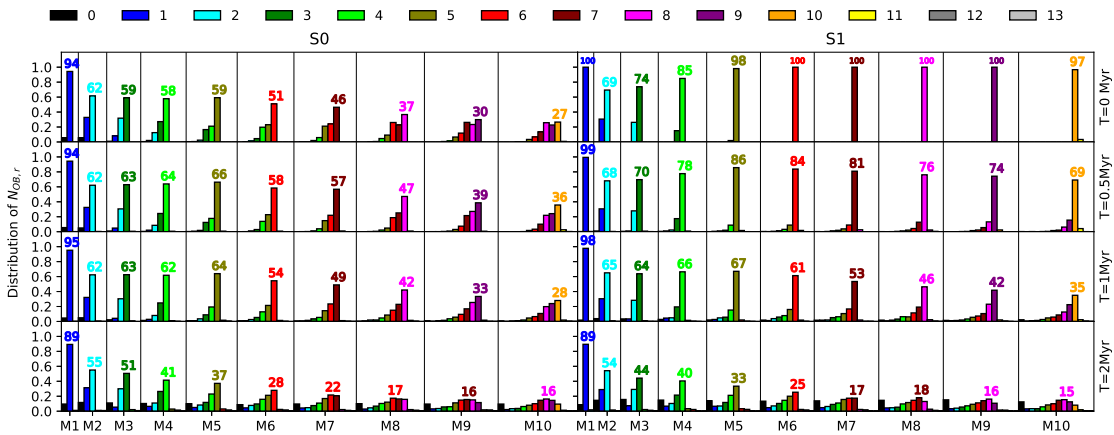}
 \caption{The distribution of the remaining number of OB stars ($N_{\mathrm{OB,r}}$) with $R<0.5$~pc at different evolution times (Y-axis labels) for model Set 1.
   Each model is shown in one box.
   The different colors of the histograms show $N_{\mathrm{OB,r}}$.
   The maximum fractions of $N_{\mathrm{OB,r}}$ for each model are labelled with their values in $\%$.
   The columns show the initially non-segregated models (left) and fully-segregated models (right).
   The rows show different evolution times.
 }
 \label{fig:set1T}
\end{figure*}


First, we investigate in general what is the chance of all OB stars being ejected within the first $2$~Myr for different cluster masses.
In Fig.~\ref{fig:set1T}, we draw the distribution of the OB star numbers, $N_{\mathrm{OB,r}}$, within $0.5$~pc to the cluster center at $0$, $0.5$, $1.0$ and $2.0$~Myr.
Here we choose $0.5$~pc as the assumed minimum criterion for the OB ionization radius.
With the optimal sampling from the IMF and the initial $M_{\mathrm{ecl}}$ variation of $\sqrt{\langle M_{\mathrm{ecl}}\rangle}$, the initial OB star numbers, $N_{\mathrm{OB,0}}$, are almost the same among all sub-models of each $M_{\mathrm{ecl}}$, except the low-mass cases ($200-400$~$M_{\odot}$).
$N_{\mathrm{OB,0}}$ scales approximately linearly with $M_{\mathrm{ecl}}$.
For each model, the maximum fraction of $N_{\mathrm{OB,r}}$ are labelled with their values in the figure.
Hereafter we denote this fraction as $f_{\mathrm{max}}(N_{\mathrm{OB,r}})$.
For the non-segregated (S0) models, not all OB stars are initially inside $0.5$~pc, while the maximum fractions of $N_{\mathrm{OB,0}}$ are the same as in the S1 models.

At $0.5$~Myr, $N_{\mathrm{OB,r}}$ in the S0 models tends to increase due to mass segregation.
In the S1 models, $f_{\mathrm{max}}(N_{\mathrm{OB,r}})$ decreases right away due to the ejection of OB stars via few-body interactions.
The decrease of the fraction is stronger for larger $M_{\mathrm{ecl}}$ because of the denser core.

At $1.0$~Myr, $f_{\mathrm{max}}(N_{\mathrm{OB,r}})$ begins to decrease for the S0 models.
For the S1 models, the cases in which all OB stars are ejected appear in all $M_{\mathrm{ecl}}$ but the fraction is very small ($<5\%$).

At $2.0$~Myr, both S0 and S1 models have significant drops of $f_{\mathrm{max}}(N_{\mathrm{OB,r}})$.
The distribution of $N_{\mathrm{OB,r}}$ becomes flat.
The fraction of no-OB-stars within $0.5$~pc also significantly increases to about $10\%$ for all $M_{\mathrm{ecl}}$.
The S1 models have a slightly larger fraction of no-OB-stars within $0.5$~pc.

However, the $10\%$ chance of no-OB-stars at $2$~Myr is still quite low to allow the ejection scenario to work.
According to this scenario, complete ejection occurred twice in a row, which means the current models only allow $1\%$ possibility.
This result indicates that with the initial configurations of the Set 1, it is very unlikely to form three stellar sequences with the ejection scenario.


\subsection{M4-S1 models with different initial OB star density}
\label{sec:res:set2}

Although the results of Set 1 indicate the ejection scenario to be unlikely, it does not mean it is ruled out, since we apply  strict constraints on the initial conditions.
The faster decrease of $N_{\mathrm{OB,r}}$ in the S1 models suggests that if the OB stars are initially distributed close to the center, the chances of OB ejection via few-body interactions is higher within $2$~Myr (Fig.~\ref{fig:set1T}).
On the other hand, this process tends to be faster when the OB stars are in a denser environment.
Thus, if the cluster is initially fully-segregated and has a higher central density, the chance of no-OB-stars remaining inside the radiation-influence-radius may be higher.
Consider the case of the ONC, we choose the M4-S1 model as the reference to create a new set of models for further investigation.
The M4-S1 model has $N_{\mathrm{OB,0}}=3-4$ in the cluster core, where a three/four body system can easily form and decay dynamically by expelling its members.

\begin{figure}
 \includegraphics[width=1.0\columnwidth]{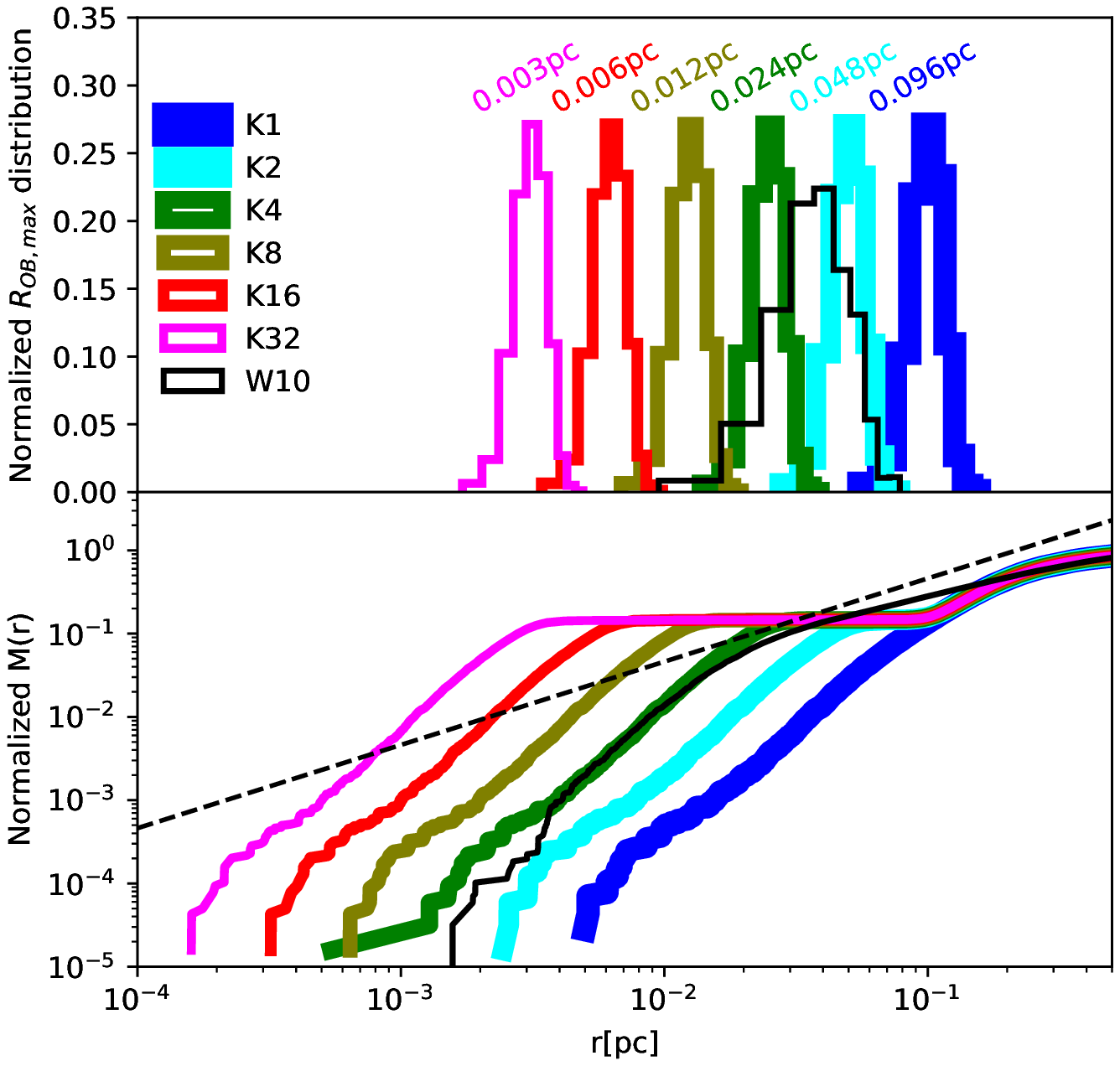}
 \caption{Upper panel: the logarithmic distribution of the maximum distance to the cluster center among all OB stars/binaries, $R_{\mathrm{OB,max}}$, for model set 2.
   The peaks of the distributions are labelled with the corresponding $R_{\mathrm{0B,max}}$.
   Bottom panel: the normalized cumulative mass distribution $M(r)$ along the radial direction by collection of all sub-models for each model in Set 2.
   The dashed black line is the singular isothermal distribution $M(r)\propto r$.
 }
 \label{fig:rdist}
\end{figure}

Firstly, we investigate the spacial distribution of the OB stars in the M4-S1 model.
In the upper panel of Fig.~\ref{fig:rdist}, we check the maximum distance of the OB binary (including OB stars as members) center-of-mass positions to the cluster center, $R_{\mathrm{OB,max}}$, of all sub-models in M4-S1 and draw the distribution (thick blue histogram).
The peak fraction occurs at $R_{\mathrm{OB,max}} \approx 0.1$~pc.
Optimal sampling from the IMF results in total masses of OB stars $M_{\mathrm{OB}}=54$~M$_\odot$ for $M_{\mathrm{ecl}}=400$~M$_\odot$.
We define the crossing timescale of an OB system as:
\begin{equation}
  T_{\mathrm{cr}}= \sqrt{ \frac{(2 R_{\mathrm{OB,h}})^3}{G M_{\mathrm{OB}}} },
  \label{eq:tcr}
\end{equation}
As the OB systems have a small number of members, the scatter of the half-mass radii of the OB systems, $R_{\mathrm{OB,h}}$, is large.
Since we focus on the case of $N_{\mathrm{OB,r}}=0$, where the most distant OB stars have a significant impact on the timescale, 
we use the $R_{\mathrm{OB,max}}$ at the peak from Fig.~\ref{fig:rdist} instead of $R_{\mathrm{OB,h}}$.
For the M4-S1 model, $T_{\mathrm{cr}}\approx 0.17$~Myr.
The complete decay of OB systems needs a few hundred $T_{\mathrm{cr}}$ \citep{Sterzik1998}.
This suggests that a significant decay cannot be expected within $1$~Myr in our models.
To make sure most of the decay can happen within $1$~Myr, a much smaller $R_{\mathrm{OB,max}}$ is necessary.

To create a denser core, we can either shrink $R_{\mathrm{h,0}}$ or change the density profile of the cluster.
However, a significant shrinking of $R_{\mathrm{h,0}}$ would be inconsistent with the observed data of embedded star clusters \citep{Marks2012}.
Thus, it is better to consider the change of the density profile.

A simple modification is to shrink the orbit size of all OB stars/binaries in the cluster by a factor of $\kappa$ according to Eq.~\ref{eq:scale}, which results in the Model Set 2.
The upper panel of Fig.~\ref{fig:rdist} shows the distribution of $R_{\mathrm{OB,max}}$ with different $\kappa$ in Set 2.
The corresponding $R_{\mathrm{OB,max}}$ values at the peak are shown (also in Table~\ref{tab:mod2}).
Table~\ref{tab:mod2} lists the $T_{\mathrm{cr}}$ and the OB star local stellar density, $\rho_{\mathrm{OB}}$, of scaling models in the Model set 2.
$\rho_{\mathrm{OB}}$ is calculated according to
\begin{equation}
  \rho_{\mathrm{OB}} = \frac{3 M_{\mathrm{OB}}}{ 4 \pi R_{\mathrm{OB,max}}^3},
  \label{eq:rho}
\end{equation}
where $R_{\mathrm{OB,max}}$ takes the same value as in the calculation of $T_{\mathrm{cr}}$.
With the smallest $\kappa=1/32$ (K32), the peak of the distribution is at $R_{\mathrm{OB,max}}\approx 3.1 \times 10^{-3}$~pc and $T_{\mathrm{cr}}\approx 9.4\times10^{-4}$~Myr with $\rho_{\mathrm{OB}}\approx 4.8 \times 10^8 $M$_\odot/$pc$^3$.
A much faster decay is expected.

The lower panel of Fig.~\ref{fig:rdist} describes the normalized cumulative mass distribution $M(r)$ along the radial direction of all sub-models of each $\kappa$.
The scaled models result in an over-density in the core region.
The singular isothermal distribution is shown in Fig~\ref{fig:rdist} for comparison.
When $\kappa<1/4$, the over-density exceeds the singular isothermal distribution.
To have a self-consistent density distribution model, we also generate a King model with $W_{\mathrm{0}}=10$.
As discussed in Section~\ref{sec:method}, above $W_{\mathrm{0}}=10$, there is no significant difference of the King models for $M_{\mathrm{ecl}}=400$~$M_\odot$.
The core density and $R_{\mathrm{OB,max}}$ distribution of the W10 model are close to the case of the K4 model.


\begin{figure}
 \includegraphics[width=1.0\columnwidth]{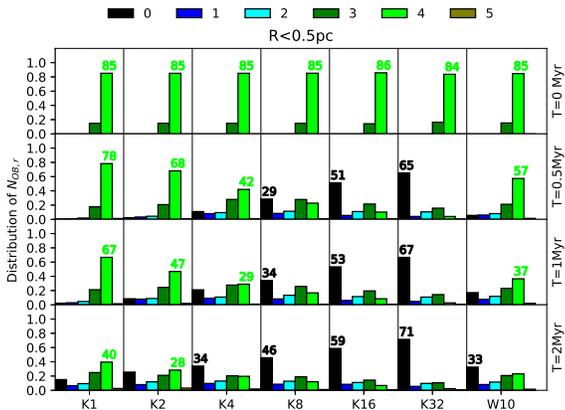}
 \caption{Similar as Fig.~\ref{fig:set1T},  the distribution of $N_{\mathrm{OB,r}}$ within $R<0.5$~pc at different evolution times for model Set 2 (fully-segregated $M_{\mathrm{ecl}}=400$~M$_\odot$ model with different orbital scaling of OB stars).
   Along one row, each box shows a different model scaling factor $\kappa$.
   The last column shows the King model with $W_{\mathrm 0}=10$.
 }
 \label{fig:modT}
\end{figure}

Fig.~\ref{fig:modT} shows the distribution of $N_{\mathrm{OB,r}}$ within $0.5$~pc of model Set 2.
Indeed, the more significant scaling results in a faster decrease of $N_{\mathrm{OB,r}}$.
Especially with $\kappa\le 1/8$, the cases of no-OB-stars remaining within $0.5$~pc become the peak fractions after $0.5$~Myr ($50\%$ in K16 and $65\%$ in K32).
At $1$~Myr, there is no significant change of $N_{\mathrm{OB,r}}$ for $\kappa \le 1/8$.
This indicates that most of the decays probably already finish within $0.5$~Myr.
The higher fraction of no-OB-stars in K32 suggests the denser OB core also result in more energetic ejections.

\begin{figure}
 \vspace{22pt}
 \includegraphics[width=1.0\columnwidth]{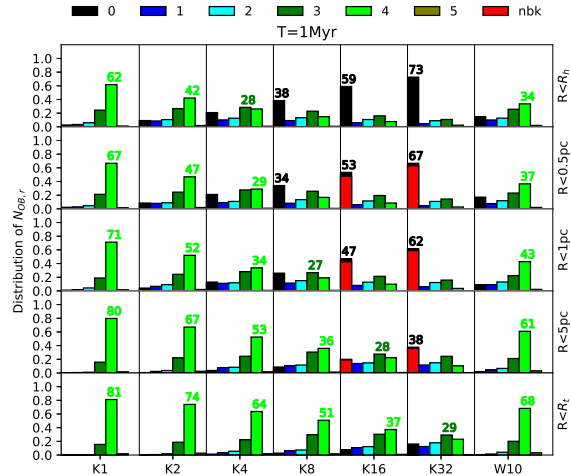}
 \caption{Similar as Fig.~\ref{fig:set1T} and \ref{fig:modT},
   the distribution of $N_{\mathrm{OB,r}}$ at $1$~Myr depending on different distance criterion for Model set 2.
   For the models K16 and K32 with $R<0.5,1,5$~pc, the red color (nbk) show the no-OB-star case in which no OB stars come back within $1$~Myr ($f_{\mathrm{nbk}} + f_{\mathrm{bk,\Delta t>1}}$ in Table~\ref{tab:fac}).
 }
 \label{fig:modR}
\end{figure}

Because the $0.5$~pc distance criterion may not be sufficient enough to avoid the radiation influence of the remaining OB stars,
we consider a different distance criterion in Fig.~\ref{fig:modR}, which is a similar plot as Fig.~\ref{fig:modT} but fixed at $T=1$~Myr.
Between $0.5$-$1$~pc, the results are not very different.
But for $R<5$~pc, the no-OB-star cases are not the dominating ones any more, except in the K32 models ($40\%$).
The chance of all OB stars being ejected outside of the tidal radius is very low for all models.

The results for the King model W10 are also shown together in Fig.~\ref{fig:modT} and \ref{fig:modR}.
As we see in Fig.~\ref{fig:rdist}, the mass distribution of the W10 model is close to the K4 model.
Their results for $N_{\mathrm{OB,r}}$ are also very similar.
This suggests that the ejection of OB stars is dominated by the OB density, the profile of the outer region has no significant impact.

\begin{table}
  \centering
  \caption{The fraction of no-OB-stars within different radius limits $R_{\mathrm{cut}}$.
    The results of the K16 and K32 models are shown.
    After OB stars are ejected out of $R_{\mathrm{cut}}$ before $1$~Myr, some may come back within the time interval $\Delta t$.
    Here three different fractions are shown:
    $f_{\mathrm{nbk}}$: the fraction of no-OB-stars with $R<R_{\mathrm{cut}}$ and no OB stars coming back within $2$~Myr.
    $f_{\mathrm{bk,\Delta t>1}}$: the fraction of no-OB-stars with $R<R_{\mathrm{cut}}$ and some coming back with $\Delta t>1$~Myr;
    $f_{\mathrm{bk,\Delta t\le1}}$: the fraction of no-OB-stars with $R<R_{\mathrm{cut}}$ and some coming back within $\Delta t\le1$~Myr;
  }
  \label{tab:fac}
  \begin{tabular}{@{}llllll@{}}
    \hline
    Model & $R_{\mathrm{cut}}[pc]$ & $f_{\mathrm{nbk}}$ & $f_{\mathrm{bk,\Delta t>1}}$ & $f_{\mathrm{bk,\Delta t\le 1}}$ \\
    \hline
    K16 & $0.5$ & $44.18\%$ & $3.59\%$ & $10.82\%$ \\
    K16 & $1.0$ & $41.12\%$ & $1.29\%$ & $8.06\%$ \\
    K16 & $5.0$ & $18.59\%$ & $0.47\%$ & $0.24\%$ \\
    K32 & $0.5$ & $60.76\%$ & $2.19\%$ & $8.26\%$ \\
    K32 & $1.0$ & $57.73\%$ & $1.20\%$ & $5.93\%$ \\
    K32 & $5.0$ & $35.29\%$ & $0.49\%$ & $1.41\%$ \\
    \hline
  \end{tabular}
\end{table}

Fig.~\ref{fig:modT} shows the distribution of $N_{\mathrm{OB,r}}$ measured at four different times.
However, an OB star ejected from a weak decay may come back to the center after a time interval $\Delta t$.
If $\Delta t$ is too short (much less than $1$~Myr), the gas in-fall may not be sufficiently fast to trigger new star formation.
In order to investigate this, for each sub-model in K16 and K32 models, we measure the starting time when all OB stars are ejected out of the radius criterion $R_{\mathrm{cut}}$.
If any OB star comes back into $R_{\mathrm{cut}}$ later, we register the returning time and calculate the time interval ($\Delta t$).
In Table~\ref{tab:fac}, we quantify the fractions of no-OB-stars (occuring before $1$~Myr) with no-come-back within $2$~Myr ($f_{\mathrm{nbk}}$), with come-back and $\Delta t>1$~Myr ($f_{\mathrm{bk,\Delta t>1}}$), with come-back and $\Delta t\le 1$~Myr ($f_{\mathrm{bk,\Delta t\le 1}}$).
The $f_{\mathrm{bk,\Delta t \le 1}}$ are $5$-$10\%$ for $R_{\mathrm{cut}}=0.5~$pc and $1.0$~pc.
The summation of $f_{\mathrm{bk,\Delta t>1}}$ and $f_{\mathrm{nbk}}$ are still significant for $R<0.5$~pc ($>47\%$) and $R<1.0$~pc ($>40\%$).
Come-back thus has a very weak influence on the cases of $R<5.0$~pc, since $f_{\mathrm{bk,\Delta t \le 1}}$ are only about $1\%$.
This is clear from Fig.~\ref{fig:modR}, where the red color shows the cases $f_{\mathrm{bk,\Delta t>1}}+f_{\mathrm{nbk}}$ from Table~\ref{tab:fac}.
Thus, the come-back of OB stars does not significantly change the conclusion that no-OB-stars is the dominating case after $1$~Myr for $R<0.5$~pc and $R<1.0$~pc in the K16 and K32 models.

\begin{figure}
 \includegraphics[width=1.0\columnwidth]{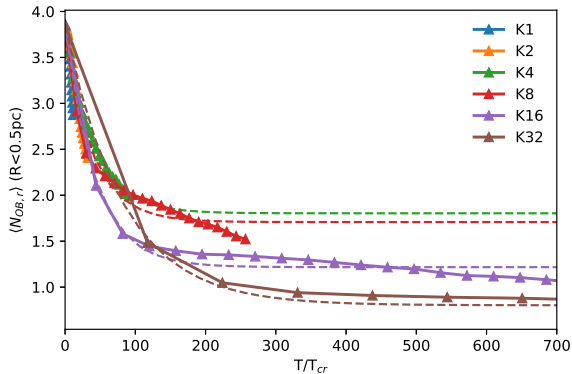}
 \caption{The evolution of average $N_{OB,r}$ of all sub-models of Model set 2. The time is scaled by the crossing time of OB stars $T_{\mathrm{cr}}$ for each model.
   The exponential decay function fittings are shown as dashed curve.
 }
 \label{fig:Nt}
\end{figure}

\begin{table*}
  \centering
  \caption{OB stellar system parameters for models in Set 2: the respective $R_{\mathrm{OB,max}}$ values at the peak in Fig.~\ref{fig:rdist}; the crossing time of the OB stellar cores (Eq.~\ref{eq:tcr}), $T_{\mathrm{cr}}$; The OB star local stellar density (Eq.~\ref{eq:rho}), $\rho_{\mathrm{OB}}$; The fitting parameters ($N_{\mathrm 0}$, $\alpha$ and $N_{\mathrm r}$) in Eq.~\ref{eq:exp} for Fig.~\ref{fig:Nt}}.
  \label{tab:mod2}
  \begin{tabular}{@{}lllllll@{}}
    \hline
    Model & $R_{\mathrm{OB,max}}$ [pc] & $T_{\mathrm{cr}}$ [Myr] & $\rho_{\mathrm{OB}}$ [M$_\odot$/pc$^3$] & $N_{\mathrm 0}$ & $\tau$ & $N_{\mathrm r}$\\
    \hline    
    K1 &  $ 9.56\times 10^{-2} $ & $ 1.70\times 10^{-1} $ & $ 1.5\times 10^{4} $ &  &  & \\
    K2 &  $ 4.78\times 10^{-2} $ & $ 6.00\times 10^{-2} $ & $ 1.2\times 10^{5} $ &  &  & \\
    K4 &  $ 2.39\times 10^{-2} $ & $ 2.12\times 10^{-2} $ & $ 9.4\times 10^{5} $ & $ 2.0134 \pm  0.0328$ & $ 40.8644 \pm 1.9338$ & $1.8036 \pm 0.0375$ \\
    K8 &  $ 1.20\times 10^{-2} $ & $ 7.50\times 10^{-3} $ & $ 7.5\times 10^{6} $ & $ 2.0009 \pm  0.1098$ & $ 40.2000 \pm 4.5063$ & $1.7079 \pm 0.0419$ \\
    K16 & $ 5.98\times 10^{-3} $ & $ 2.65\times 10^{-3} $ & $   6\times 10^{7} $ & $ 2.6203 \pm  0.0982$ & $ 43.4018 \pm 3.6168$ & $1.2174 \pm 0.0248$ \\
    K32 & $ 2.99\times 10^{-3} $ & $ 9.38\times 10^{-4} $ & $ 4.8\times 10^{8} $ & $ 3.0443 \pm  0.0698$ & $ 82.2774 \pm 5.0612$ & $0.8022 \pm 0.0168$ \\
    \hline
  \end{tabular}
\end{table*}

The above results of model Set 2 suggest that when the OB stars are initially distributed in the cluster core with a very high density, the ejection scenario can work with a reasonable probability.
Fig.~\ref{fig:Nt} displays the averaged time evolution of $N_{\mathrm{OB,r}}$ as a function of time in units of the initial crossing time (Eq.~\ref{eq:tcr}), $T_{\mathrm{cr}}$, for each model.
In the beginning, the $N_{\mathrm{OB,r}}$ has a sharp drop.
The decay data overlap, which suggests the decay time strongly depend on the $T_{\mathrm{cr}}$.
The sharp drop finishes after about $200~T_{\mathrm{cr}}$ and results in $N_{\mathrm{OB,r}}=1,2$.
This indicates the fast ejection of one OB binary.

After the sharp drop, the decrease of $N_{\mathrm{OB,r}}$ becomes much slower with a different slope.
This is clearly seen for the K8 and K16 models.
The $N_{\mathrm{OB,r}}$ at the changeover points are between $1.5$ to $2.0$.
When a multiple system containing $3-4$ stars decays, two binaries or one binary with one single are ejected into opposite directions with different velocities.
The velocity depends on the masses of the binary and the single star.
Thus two different slopes of $N_{\mathrm{OB,r}}$ are expected.

On the other hand, denser OB systems also result in a smaller final $N_{\mathrm{OB,r}}$ after about $200~T_{\mathrm{cr}}$.
This is probably due to the more energetic ejections produced by a more violent environment.
Especially in the K32 model, $N_{\mathrm{OB,r}}$ is less than one after about $200~T_{\mathrm{cr}}$.

We use the exponential decay function
\begin{equation}
  N_{\mathrm{OB,r}}\left(\frac{T}{T_{\mathrm{cr}}}\right) = N_{\mathrm 0} e^{-\frac{T}{\tau T_{\mathrm{cr}}}} + N_{\mathrm r}
  \label{eq:exp}
\end{equation}
to fit the results in Fig.~\ref{fig:Nt} and obtain a good description of the $N_{\mathrm{OB,r}}$ evolution for the drop.
The fitting parameters are listed in Table~\ref{tab:mod2}.
The first two models (K1 and K2) have not yet finished the drop, thus the fitting cannot work.
Interestingly, the exponential power coefficients $\tau$ have roughly the same value for K4-K16 ($\tau \approx 40$).
The K32 model has $\tau \approx 82$.
But K32 has only one data point during the drop, thus the fitting result can be poor, which may explain the difference.

The remaining number of OB stars after the exponential decay, $N_{\mathrm r}$, decreases when the OB density increases.
For the K32 model, $N_{\mathrm r} \approx 0.8$, which represents the significant fraction of no-OB-star cases shown in Fig.~\ref{fig:modT}.
If we extend the OB density to higher than K32, $N_{\mathrm r}$ decreases even further.
In the extreme case $N_{\mathrm r}=0$ (which represents that OB stars are always ejected completely finally) and $N_{\mathrm 0}=4$ (initially $4$ OB stars), after $150~T_{\mathrm{cr}}$ (with $\tau=40$), the $N_{\mathrm{OB,r}}$ drops to $0.1$.

\section{Discussion}
\label{sec:discussion}

The results of model Set 1 (Fig.~\ref{fig:set1T}) indicate that if the initial conditions are given by a Plummer model, it is very difficult to eject all OB stars within $1$~Myr beyond $0.5$~pc.
Then in model Set 2, we modify the density profile of the M4-S1 model that represents the case of the ONC, to see how the OB star ejection depends on the local density.
It is found that with a high central density (K16 and K32 models), the possibility to eject all OB stars out beyond $0.5$~pc within $1$~Myr is high ($50-67\%$; Fig.~\ref{fig:modT} and \ref{fig:Nt}).
The chance that any OB star comes back within $1$~Myr is low (Fig.~\ref{fig:modR} and Table~\ref{tab:fac}).
This indicates that whether the ejection scenario works sensitively depends primarily on the OB star initial spatial distribution and local density.

The K16 and K32 models have an overdensity in the core compared to the King model W10 and the singular isothermal profile (Fig.~\ref{fig:rdist}).
The recent ALMA observation of the center of the Serpens South cloud indicates that primordial mass-segregation exists in extremely young embedded clusters \citep{Plunkett2018}.
Besides, the projected separation between pairs of $50$ millimeter sources can be as small as $0.002$~pc and the peak separation is about $0.05$~pc with a large fraction of separations below the peak.
The spacial distribution of sources is irregular with separated clumps.
This suggests the OB stars can form in a very dense environment.

If the ONC had a similar spatial distribution of protostars at about $0.2$~Myr like Serpens South, the ejection scenario seems to be credible.
The central Trapezium system \citep{Hillenbrand1997} has a present-day diameter of about $0.05$~pc, which is close to the case of the K2 model.
It is possible that the system is already dynamically evolved and was denser in the past and comparable to the cases of the K16 and K32 models.
Indeed, the Trapezium has a velocity dispersion too large for dynamical equilibrium, suggesting it is in the process of dynamically decaying \citep{Pflamm2006,Subr2012,Kroupa2018}.


\subsection{Impact of initial conditions}

However, we should also discuss the possibility to break the constraints on the initial conditions in our models.
Firstly, we apply optimal sampling from the IMF, which assumes the ONC has an IMF strictly following the canonical IMF with the resulting $m_{\mathrm{max}}-M_{\mathrm{ecl}}$ relation.
It is possible that the IMF of the ONC is abnormal.
Also with the optimal sampling, the uncertainty of the total stellar masses can result in a different number of OB stars.
In the case that the OB star number were larger than $4$ for each episode of star formation, or only $1-2$ OB star formed, it would be difficult to remove all OB stars and the ejection scenario would be unlikely.
Also if no OB stars were to form at the first and second stages, it would be difficult to understand why the star formation is not smoothly continued but is split into three stages.

Secondly, we apply the \cite{Kroupa1995b,Belloni2017} model to generate the initial binary population.
The OB binaries follow the distribution of \cite{Sana2012}.
Thus our sub-models of each $M_{\mathrm{ecl}}$ cover a wide range of mass-ratio, eccentricity and period combination.
We have checked the relation between $N_{\mathrm{OB,r}}$ and the OB binary properties and found no clear evidence showing they are correlated (not shown as plot here).
Therefore, the dynamical modification of OB binary properties is not likely to give significantly different results.

Thirdly, the \cite{Marks2012} $R_{\mathrm{h,0}}-M_{\mathrm{ecl}}$ relation is assumed to constrain the $R_{\mathrm{h,0}}$ of our models.
With the same density profile,  $R_{\mathrm{h,0}}$ can have a strong impact on the results since it changes the central density.
However, the results of model Set 2 (Fig.~\ref{fig:modT} and \ref{fig:modR}) indicate that to allow the ejection scenario to work, the OB stars should be in a $10$ times denser environment, which means $R_{\mathrm{h,0}}$ should be shrunk to about $0.02$~pc.
This is too dense given the observed properties of embedded clusters. 
The disruption of binaries in such a dense cluster will result in a significantly lower binary fraction compared to the observational data of the ONC \citep[e.g][]{Kroupa2001b,Duchene2018}.

Fourthly, the gas dynamics is simply ignored in this work.
If a gas potential exists, the initial virial ratio of the stellar system can be significantly different.
Especially, the averaged velocity dispersion of the stars can be enhanced (Section 6.4 in \citealp{Kroupa1995a}).
It may accelerate the mass segregation of the OB stars in the S0 models.
But in the S1 models, all OB stars are already primordially segregated to the cluster center, which is consistent with observational indications \citep{Kirk2011,Plunkett2018}.
The gas potential influence is weak on the decay timescale of OB few-body systems, unless the gas has a cuspy density distribution in the cluster core.
On the other hand, the existence of gas enlarges the escape velocities from the clusters, which makes the OB runaways more difficult to be ejected out of the cluster.
Especially if the decay is weak, the OB stars may come back in a short timescale.
However, the ionizing stars lead to a rapid evolution of the gas density, such that accurate treatment of complex gas-plus-star system is not readily possible.

\begin{table*}
  \centering
  \caption{The frequency of initial OB star ($m\ge 8$~M$_\odot$) numbers $N_{\mathrm{OB,0}}=3,4$ by randomly sampling from the Kroupa (2001) IMF $10000$ times.}
  \label{tab:samp}
  \begin{tabular}{@{}lllllllllll@{}}
    \hline
    Model & M1 & M2 & M3 & M4 & M5 & M6 & M7 & M8 & M9 & M10 \\
    Frequency($\%$) & 3.93 & 29.59 & 47.62 & 43.48 & 28.30 & 15.79 & 7.32 & 3.25 & 1.27 & 0.51 \\
    \hline
  \end{tabular}
\end{table*}

\subsection{Stochastic or self-regulated}
\label{sec:dis:2}

The discussion above indicates that the conditions, especially the number of OB stars and their density, to allow the ejection scenario to work, are strongly restricted.
The number of OB stars sensitively depends on the IMF.
We did a test with random sampling from the \cite{Kroupa2001} IMF $10000$ times for each $M_{\mathrm {ecl}}$ in model Set 1 to obtain the frequency of the cases with initial OB star numbers $N_{\mathrm{OB,0}}=3,4$.
The result is shown in Table~\ref{tab:samp}.
The M3 and M4 models have, in about half of the cases, $N_{\mathrm{OB,0}}=3,4$.
For the other models, the frequencies are much lower.
The test repeated $10000$ times gives a Poisson scatter of $100$.
Thus there is a significant scatter for the models with frequency below $10\%$.
This result indicates that if the random sampling from the IMF within star clusters were preferred in nature, the chance to obtain twice $N_{\mathrm{OB,0}}=3,4$ for two consecutive star formation events would be rather low (upper limit about $25\%$).

However, if star formation within an embedded clusters were stochastic, the observational results of primordial mass-segregation by \cite{Plunkett2018} would be difficult to understand.
The $m_{\mathrm{max}}-M_{\mathrm{ecl}}$ relation would also not exist \citep{Weidner2013,Weidner2014}.
The primordial mass-segregation suggests that the star formation process depends on its local environment and is self-regulated.
This is consistent with being an optimally sampled distribution function rather a probability density distribution function and the $m_{\mathrm{max}}-M_{\mathrm{ecl}}$ relation.
Therefore, if the ejection scenario is the true reason for the multi-sequence star formation in the ONC, it also suggests that the star formation should be a self-regulated process, and is consistent with observed primordial mass-segregation.

\subsection{Future observational constraints}
The chance of multi-sequence star formation in young star clusters driven by the ejection scenario is only limited to a small region of $M_{\mathrm{ecl}}$ (around $400$~M$_\odot$), which allows only $3$-$4$ OB stars to be formed.
Future observations can test the ejection scenario by searching whether multi-sequence with short age gaps ($<2$~Myr) appear in lower- or higher-mass star clusters.
This is further discussed in \cite{Kroupa2018}.




\section{Conclusion}
\label{sec:conclusion}

In this work, we have performed detailed $N$-body models to investigate the formation scenario for the three stellar sequences observed in the ONC \citep{Beccari2017} proposed by \citep{Kroupa2018}.
\cite{Kroupa2018} scenario suggests that the OB stars provide radiation feedback to quench the star formation within a short timescale (about $1$~Myr).
But if all OB stars are ejected out of the radiation influence radius via few-body dynamical decay, gas in-fall can switch on star formation again.
If these processes occur twice, the observed three sequences can be well explained.

Our standard $N$-body models (Set 1) with the Plummer profile and observationally constrained primordial binary properties and half-mass radii, suggest that the chance of no-OB-stars within $0.5$~pc after $1$~Myr is very low ($<10\%$).
However, by breaking the assumption of a Plummer profile, we find that a fully mass-segregated $400$~M$_\odot$ cluster with $3-4$ OB stars and an about $10$ times denser OB system than normally given by the Plummer model in the cluster center, can significantly enhance OB ejections making the possibility of the ejection scenario more likely.
The $N$-body calculations demonstrate that in the great majority of cases ejected OB stars do not return to the inner $1$~pc region of the cluster within $< 1$~Myr.
If this scenario is the channel to form the multi-sequence, it also indicates that the star formation should be self-regulated and close to optimal or deterministic.

The exponential decay of averaged $N_{\mathrm{OB,r}}$ in Fig.~\ref{fig:Nt} suggests that the OB decay timescale is about $40~T_{\mathrm{cr}}$.
In an extremely dense environment, $N_{\mathrm{OB,r}}$ may reach $0.1$ after $150 T_{\mathrm{cr}}$.
Thus, if the birth density of OB stars in embedded clusters is very high with maximum separation $<5\times10^{-3}$~pc (OB stellar density $> 10^8$~M$_\odot$/pc$^3$), the ejection scenario can be an important channel to form Myr age spread multi stellar-sequences in young star clusters with $3-4$ OB stars forming in each sequence.

In addition, we also improve the star cluster initial model generator tool \textsc{mcluster} by updating the primordial binary properties based on the new observational constraints, accelerating the performance and fixing several bugs (see Appendix).

\section*{Acknowledgments}
L.W. thanks the Alexander von Humboldt Foundation for funding this research.

\appendix
 
\section[]{New implementation of \textsc{mcluster}}

We have updated the star cluster initial model generator code \textsc{mcluster}, which was originally developed by \cite{Kupper2011}.
The new version includes the updated binary properties constrained by the recent observational data, accelerates the code via GPU parallelization for large particle numbers, adds the support for the \textsc{nbody6++gpu} code and fixes a few bugs.
Here we describe the details of these new implementations.

\subsection{Primordial binary property}
\subsubsection{Pre-main-sequence eigenevolution}
In \cite{Belloni2017}, the pre-main-sequence eigenevolution of \cite{Kroupa1995b} is modified to improve the representation of parameter correlations with period for different primary star masses.
Eq.8-11 in \cite{Belloni2017} is implemented in the new version of \textsc{mcluster}.
The new pre-main-sequence eigenevolution procedure is switched on by default (the variable eigen=2 in the source code).

\subsubsection{Binaries with high-mass stars}
For the high-mass stars ($>5$~M$_\odot$), the primordial binary properties are updated based on \cite{Sana2012,Oh2015,Belloni2017} (see our Section~\ref{sec:init}).
Users of \textsc{mcluster} can choose this distribution by option ``-p 3''.

\subsection{GPU acceleration for large N}
The scaling from astronomical units to $N$-body units (H\'enon units) requires the calculation of the total potential energy.
This is an O($N^2$) calculation where $N$ is the total number of particles.
When $N>10^5$, this calculation by \textsc{mcluster} is very time consuming.
The GPU potential library from \textsc{nbody6-gpu} \citep{Nitadori2012} is imported to accelerate the potential calculation.
This is significantly more efficient for the case of $N=10^6$.

\subsection{Support for \textsc{nbody6++gpu}}
The interface to the up-to-date version of \textsc{nbody6++gpu} \citep{Wang2015} (Apr 27, 2018) is implemented.
Users of \textsc{mcluster} can use option '-C 5' to generate initial conditions for it.

\subsection{Bug fixes}
\subsubsection{Scaling issue}
In the original \textsc{mcluster} code, the gravitational constant has the value:
\begin{equation}
  G = 0.0043009211 pc M_\odot^{-1} (km/s)^2.
\end{equation}
However, this is inconsistent with the gravitational constant used in the \textsc{nbody6} series of codes, which comes from the IAU 2009 System of Astronomical Constants \citep{Luzum2011}.
This results in an about $4.5\%$ systematical error in initial stellar masses.
The velocity scaling of stars is also affected, especially, the binary orbit parameters are inconsistent with the models used.
The correct gravitational constant (now implemented in the new version of \textsc{mcluster}) is:
\begin{equation}
  G = 0.00449850214 pc M_\odot^{-1} (km/s)^2.
\end{equation}

Besides, when pre-main-sequence eigenevolution is used, some of the secondary stars in binaries gain masses.
However, the original \textsc{mcluster} does not include the corresponding correction of the total mass, which results in an additional systematical error of the total masses (about $5\%$).
Again, the velocity scaling is also influenced.

\textsc{mcluster} first generates the orbits of stars from the density profile, then creates the binaries.
The pre-main-sequence eigenevolution, which increases the masses of secondaries with tight orbits, is done after the binary generation.
Thus, the virial ratio is changed due to the larger mass of the system.
However, there is no re-scaling to correct the virial ratio after that.
The users can observe that the final virial ratio value shown in \textsc{nbody6(++)} is different from the initial input value for \textsc{mcluster}.

\subsubsection{King model generator}
The original \textsc{mcluster} code supports the King model generator for $W_{\mathrm 0}$ ranging from $1$ to $12$.
The upper limit of $12$ is removed in the new version with corresponding modification.

On the other hand, there is a bug in the King model generator to determine the step size of the ordinary differential equation integrator used to calculate the density profile.
Fortunately, the effect of the bug is not significant, which caused difficulty to identify it.

%


\label{lastpage}

\end{document}